\documentclass[twoside]{dis09}
\usepackage[latin1]{inputenc}
\usepackage[dvips]{graphicx,epsfig,color}
\usepackage{wrapfig,rotating}
\usepackage{amssymb,amsmath,array}
\newcommand{\be}{\begin{equation}}
\newcommand{\ee}{\end{equation}}
\newcommand{\bea}{\begin{eqnarray}}
\newcommand{\eea}{\end{eqnarray}}
\newcommand{\bfS}{\mbox{\boldmath $S$}}

\newcommand{\bfp}{\mbox{\boldmath $p$}}
\newcommand{\bfk}{\mbox{\boldmath $k$}}

\newcommand{\pup}{p^\uparrow}
\newcommand{\qup}{q^\uparrow}
\newcommand{\qdown}{q^\downarrow}
\newcommand{\pdown}{p^\downarrow}
\def\lsim{\mathrel{\rlap{\lower4pt\hbox{\hskip1pt$\sim$}}\raise1pt\hbox{$<$}}}
\def\gsim{\mathrel{\rlap{\lower4pt\hbox{\hskip1pt$\sim$}}\raise1pt\hbox{$>$}}}
\def\nostrocostruttino#1\over#2{\mathrel{\mathop{\kern 0pt \rlap
{\hbox{$#1$}}} \hbox{\kern-.135em $#2$}}}

\begin{document}
\title{Sivers and Collins Effects: from SIDIS \\ to Proton-Proton Inclusive Pion Production}

\author{
M. Anselmino$^{1,2}$, M. Boglione$^{1,2}$, U. D'Alesio$^{3,4}$, E. Leader$^{5}$,
S. Melis$^{2,6}$, F. Murgia$^{4}$, A. Prokudin$^{1,2}$
%
\vspace{.3cm}\\
%
1- Universit\`a di Torino - Dipartimento di Fisica Teorica - \\
Via P. Giuria 1, I-10125 Torino - Italy
\vspace{.1cm}\\
2- INFN - Sezione di Torino - \\ Via P. Giuria 1, I-10125 Torino - Italy
\vspace{.1cm}\\
3- Universit\`a di Cagliari - Dipartimento di Fisica - \\ C.P. 170, I-09042
Monserrato (CA) - Italy
\vspace{.1cm}\\
4- INFN - Sezione di Cagliari - \\ C.P. 170, I-09042 Monserrato (CA) - Italy
\vspace{.1cm}\\
5- Imperial College London- South Kensington Campus \\ Prince Consort Road, London SW7 2AZ - U.K.
\vspace{.1cm}\\
6- Universit\`a del Piemonte Orientale -\\ Dipartimento di Scienze e Tecnologie
Avanzate -\\ Viale T. Michel 11, I-15121 Alessandria - Italy
}

\maketitle

\begin{abstract}
We consider the Sivers, Collins and transversity functions as extracted from SIDIS
and $e^+e^-$ experimental data and investigate to what extent they might explain the
large Single Spin Asymmetries (SSA) observed in proton-proton inclusive processes.
This  phenomenological study is performed within the TMD factorization scheme.
As the  SIDIS data cover only a limited range of $x$ values
($x \lsim 0.3$), we allow for different large $x$ behaviours of the SIDIS
Sivers functions and transversity distributions. We conclude that, within the
available experimental constraints, one cannot observe any clear universality breaking effect for the Sivers functions.
\end{abstract}

We report \cite{url} on some work in progress, exploring the simple phenomenological
idea of adopting the same Sivers, Collins and transversity functions, as extracted from
SIDIS and $e^+e^-$ experimental data~\cite{Anselmino:2008sga, Anselmino:2008jk}, to evaluate the corresponding Sivers and Collins effects in proton-proton scattering, assuming a TMD factorized scheme. These effects are then compared to the RHIC
proton-proton data on SSAs at $\sqrt s =$ 200 GeV~\cite{star,brahms}. \\
The $A_{UT}^{\rm Sivers}$ transverse single spin asymmetry, measured by the
HERMES~\cite{Airapetian:2004tw,Diefenthaler:2007rj} and
COMPASS~\cite{Alexakhin:2005iw,Ageev:2006da,Martin:2007au,Alekseev:2008dn}
collaborations in $\ell N \to \ell h X$ SIDIS processes, has been analyzed according to the expression:
\be
A_{UT}^{\rm Sivers} \propto
\frac{\sum _q e^2_q \, \Delta \hat{f}_{q/p^\uparrow}(x,\bfk_\perp) \otimes
\frac{d\hat\sigma ^{\ell q \to \ell q}}{dQ^2} \otimes D_{h/q}(z,p_\perp) }
     {2\sum _q e^2_q \, f_{q/p}(x,k_\perp) \otimes
     \frac{d\hat\sigma ^{\ell q \to \ell q}}{dQ^2} \otimes D_{h/q}(z,p_\perp)}\,,
\label{sidis-sivers}
\ee
where $f_{q/p}(x,k_\perp)$ and $D_{h/q}(z,p_\perp)$ are the unpolarized
distribution and fragmentation functions, with $\bfk_\perp$ and $\bfp_\perp$
being, respectively, the transverse momentum of the quark in the proton and of
the final hadron $h$ with respect to the fragmenting quark $q$;
$\frac{d\hat\sigma ^{\ell q \to \ell q}}{dQ^2}$ is the partonic cross section corresponding to the underlying elementary process $\ell q \to \ell q$. The
numerator of this azimuthal asymmetry contains the Sivers distribution
function~\cite{Sivers},
$\Delta \hat f_{q/p^\uparrow}(x,\bfk_\perp)$, related to the number density of
unpolarized quarks 
inside a transversely polarized proton 
\be
\Delta \hat f_{q/p^\uparrow}(x,\bfk_\perp)=\hat f_{q/\pup}(x, \bfk_{\perp})
- \hat f_{q/\pdown}(x, \bfk_{\perp})
\equiv \Delta^N f_{q/\pup}\,(x, k_{\perp}) \>
\bfS_T \cdot (\hat{\bfp} \times \hat{\bfk}_{\perp }) \label{defsiv}\,.
\ee
For the purpose of our fit, we parametrize this function by factorizing the
$x$ and $k_\perp$ dependences, as follows~\cite{Anselmino:2008sga}
\be
\Delta^N  f_{q/\pup}\,(x, k_{\perp})  \propto x^{\alpha_q} (1-x)^{\beta_q}
f_{q/p}(x) \,h(k_{\perp}) \>,
\ee
where $\alpha_q$ and $\beta_q$ are free parameters which control the details
of the low-$x$ and large-$x$ behaviour of the Sivers function, for each given
flavour $q$.

%
\begin{wrapfigure}{r}{0.6\columnwidth}
\vspace*{-1cm}
\begin{center}
\hspace*{0.7cm}
\includegraphics[width=0.22\textwidth,angle=-90,bb = 20 150 490 480]{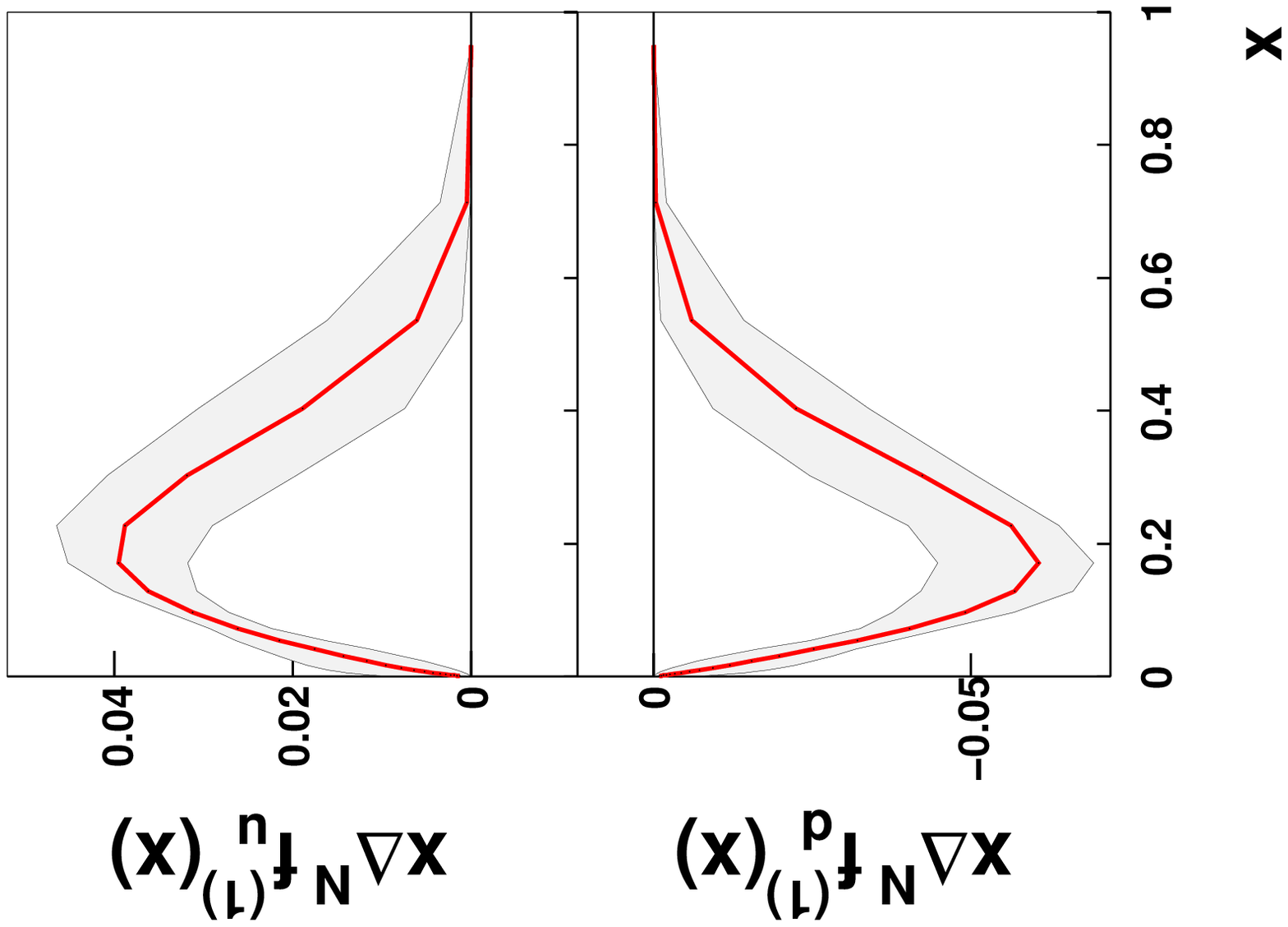}
\includegraphics[width=0.22\textwidth,angle=-90,bb = 20 150 490 480]{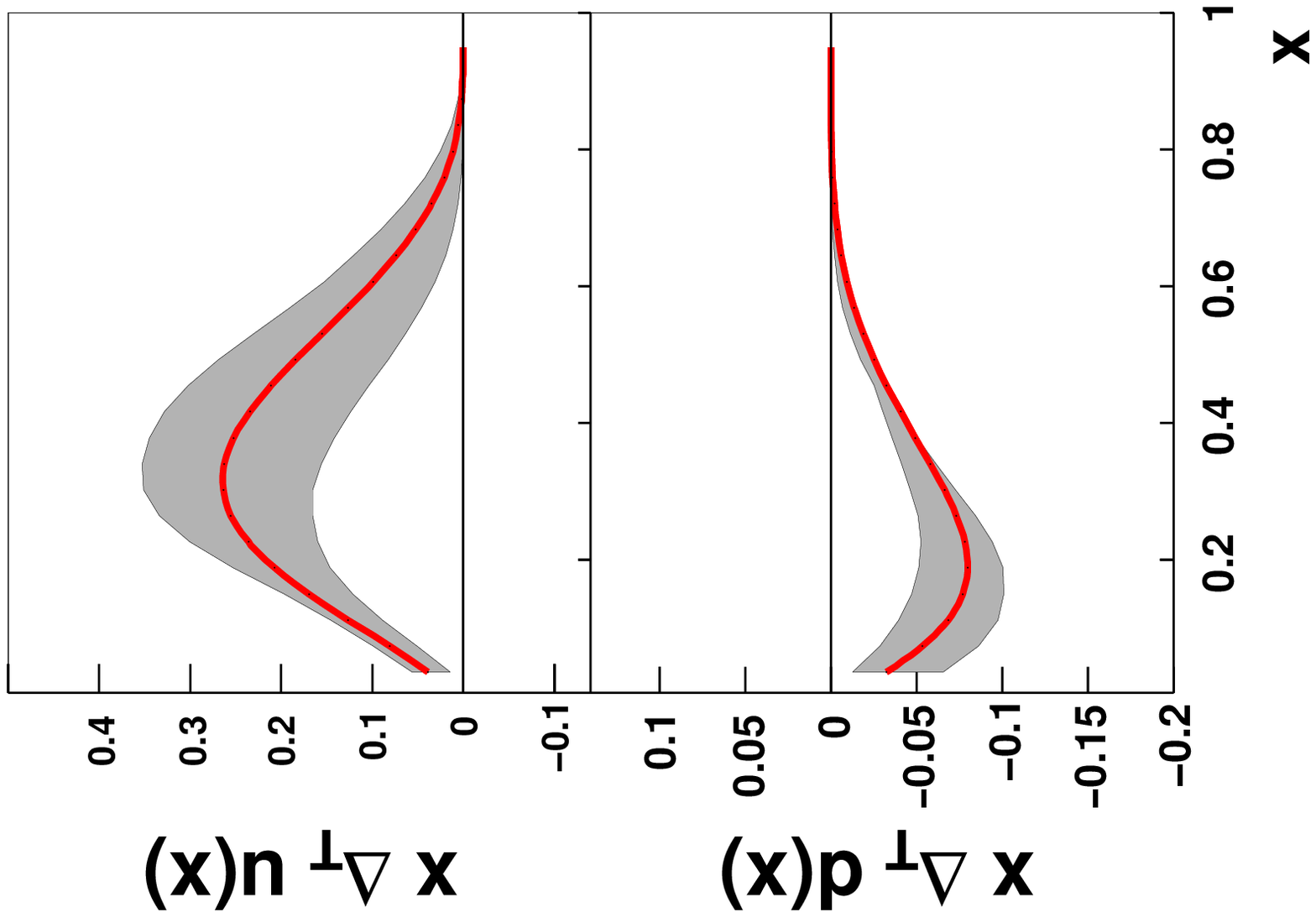}
\includegraphics[width=0.22\textwidth,angle=-90,bb = 20 150 490 480]
{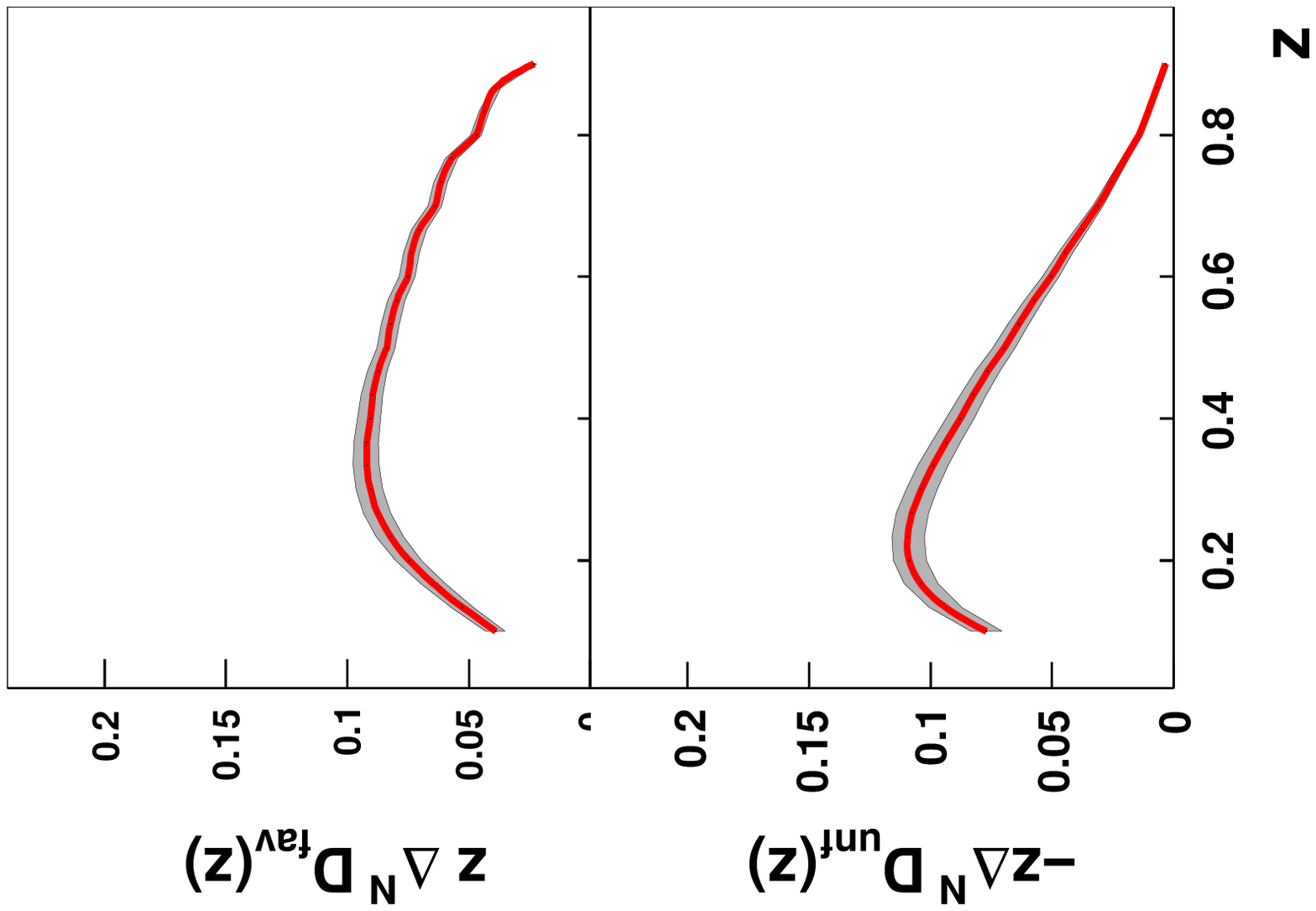}
\end{center}
\caption{The Sivers, transversity and Collins density functions as determined by
fitting SIDIS and $e^+e^-$ data \cite{Anselmino:2008sga},\cite{Anselmino:2008jk}.}
\label{Fig:density-fns}
\end{wrapfigure}
%
A fit of the available experimental data allowed us the extraction of the
Sivers functions, shown in the first panel of Fig.~\ref{Fig:density-fns}, by
using the unpolarized distribution and fragmentation function sets (with their
appropriate $Q^2$ dependence) as given in
Refs.~\cite{Gluck:1998xa} and \cite{deFlorian:2007aj} respectively.\\
The transversity and the Collins functions, which, being chirally odd, can only
contribute in pairs to physical observables, were determined in
Ref.~\cite{Anselmino:2007fs} and updated in Ref.~\cite{Anselmino:2008jk} by
performing a simultaneous fit of the Collins azimuthal asymmetry,
$A_{UT}^{\rm Collins}$, measured in SIDIS by HERMES and COMPASS, %
\be
A_{UT}^{\rm Collins} \propto
\frac{\sum _q e^2_q \,h_{1q}(x,\bfk_\perp) \otimes
\frac{d\Delta\hat\sigma ^{\ell q \to \ell q}}{dQ^2}
\otimes \Delta \hat D_{h/q^\uparrow}(z,\bfp_\perp)}
     {2\sum _q e^2_q \, f_{q/p}(x,k_\perp) \otimes
\frac{d\hat\sigma ^{\ell q \to \ell q}}{dQ^2}
\otimes D_{h/q}(z,p_\perp)}\,,
\label{sidis-collins}
\ee
and of the azimuthal correlations $A_{12}$ measured in $e^+e^-\to h_1h_2\,X$
by the BELLE collaboration~\cite{Abe:2005zx,Seidl:2008xc}.
$h_{1q}(x,\bfk_\perp)$ and $\Delta \hat D_{h/\qup}(z,\bfp_\perp)$ are the
transversity and Collins functions and
$\frac{ d\Delta\hat\sigma ^{\ell q \to \ell q}}{dQ^2}$ is the partonic spin
transfer cross section, while $A_{12}$ contains the product
of two Collins functions.
%

The Collins function is related to the number density of unpolarized hadrons
$h$ resulting from the fragmentation of a transversely polarized quark:
\be
\Delta \hat D_{h/\qup}\,(z, \bfp_\perp)
= \hat D_{h/\qup}\,(z, \bfp_\perp) -
\hat D_{h/\qdown}\,(z, \bfp_\perp)
\equiv \Delta^N D_{h/\qup}\,(q, p_{\perp}) \>
\bfS_q \cdot (\hat{\bfp}_q \times \hat{\bfp}_{\perp }) \>,
 \label{defcolnoi}
\ee
where $\bfS_q$ and ${\bfp}_q$ are, respectively, the polarization vector and
the momentum vector of the fragmenting quark $q$, while $\bfp_\perp$ is the
intrinsic transverse momentum of the produced hadron with respect to the
$\hat{\bfp}_q$ direction.
%

\noindent
As in the Sivers fit, the transversity (Collins) functions were parametrized
so that their $x$ $(z)$ and $k_\perp$ ($p_\perp$) dependences were factorized
and their low-$x$ (low-$z$) and large-$x$ (large-$z$) behaviour controlled by
the appropriate $\alpha$ and $\beta$ parameters.
The transversity and the Collins functions as determined using this strategy are
shown in Fig.~\ref{Fig:density-fns}. One can observe that while the Collins
fragmentation functions are rather well constrained thanks to the high statistics
of the BELLE data~\cite{Anselmino:2008jk}, the Sivers and the transversity
distributions are affected by much higher uncertainties. Moreover, it is
important to notice that the available SIDIS experimental data span a
relatively limited range of $x$ values ($x \lsim 0.3$): therefore, the SIDIS
data are unable to fix the parameters $\beta$ which control the large-$x$
behaviour of the transversity and of the Sivers distribution functions. As a
consequence, in our fits the $\beta$ parameters were chosen to be flavor
independent. This observation is relevant when turning to polarized
proton-proton SSAs~\cite{Anselmino:2005sh}; in fact, $pp$ experimental data from RHIC cover a range
of much larger $x$ values as compared to SIDIS data. When exploring the Sivers and the
Collins effects induced in $pp$ processes by the SIDIS extracted functions, this large
$x$ uncertainty should be taken into account.

\noindent
The actual consensus and understanding about the Collins and Sivers functions
is that while the former are expected to be universal~\cite{Collins:2004nx,Yuan:2009dw,
Gamberg:2008yt}, the latter can be process dependent.
%
%
\noindent
In particular the Sivers functions in SIDIS and Drell-Yan processes are expected to
be opposite. The situation is much less clear concerning SSAs in $AB \to hX$
processes in which the only large scale measured is the $P_T$ of the final
hadron $h$~\cite{Bacchetta:2005rm,Bomhof:2006dp,Ratcliffe:2007ye}; even the
factorization scheme with transverse momentum dependent distribution and
fragmentation functions (TMD factorization) has not been proven in such cases.
%

\noindent
We adopt here a pragmatic attitude and explore the possible values of SSAs
in $\pup p \to \pi X$ processes at moderately large $P_T$, assuming TMD factorization and
universality, that is using the same Sivers and Collins functions as extracted
from SIDIS data. A failure to reproduce the experimental results would be
a clear indication that these assumptions cannot be valid. The expression of
$A_N$ in TMD factorization is given by~\cite{Anselmino:2005sh}:
\bea
A_N &\sim& A_N^{\rm Sivers} + A_N^{\rm Collins} \nonumber \\
& \propto &
\frac{\displaystyle \sum _{a,b,c,d} 
\, \Delta \hat f_{a/p^\uparrow}(x_a,\bfk_{\perp a}) \otimes \,f_{b/p}(x_b,k_{\perp b}) \otimes \frac{d\hat\sigma ^{ab\to cd}}{dt} \otimes D_{h/c}(z,p_\perp) }
     {\displaystyle 2 \!\! \sum _{a,b,c,d} \, f_{a/p}(x_a,k_{\perp a})\otimes \, f_{b/p}(x_b,k_{\perp b}) \otimes \frac{d\hat\sigma ^{ab\to cd}}{dt} \otimes D_{h/c}(z,p_\perp)} \nonumber \\
&+&
\frac{\;\;\displaystyle \sum _{a,b,c,d}  \, h_1(x_a,\bfk_{\perp a})  \otimes \,f_{b/p}(x_b,k_{\perp b}) \otimes \frac{d\Delta\hat\sigma^{ab\to cd}}{dt} \otimes 
\Delta \hat D_{h/c^\uparrow}(z,\bfp_\perp) }
     {\displaystyle 2  \!\! \sum _{a,b,c,d} \, f_{a/p}(x_a,k_{\perp a}) \otimes \, f_{b/p}(x_b,k_{\perp b}) \otimes \frac{d\hat\sigma^{ab\to cd}}{dt} \otimes D_{h/c}(z,p_\perp)}\;,
\label{AN}
\eea
where $a$, $b$, $c$ and $d$ can be either quarks $q$ or gluons $g$, 
and all possible pQCD elementary interactions at lowest order contribute.
Notice that the Sivers and Collins effects add up in $A_N$, and cannot be
separated as it is done in SIDIS. Further contributions, proportional to the
Boer-Mulders and other TMDs, are negligible, as we have checked
numerically~\cite{Anselmino:2005sh}.

\noindent
We now apply the quark Sivers, transversity and Collins functions extracted by fitting
SIDIS and $e^+e^-$ experimental data to compute $A_N$ for $\pup p \to \pi X$
processes~\cite{Anselmino:2005sh} and compare with data~\cite{star,brahms}.
As mentioned above, the available SIDIS experimental measurements refer to a limited
range of $x$ values ($x \lsim 0.3$): therefore, the SIDIS data are unable to
fix precisely the parameters $\beta$ which control the large-$x$ behaviour
of the transversity and of the Sivers distribution functions, while the RHIC
$pp$ experimental data cover a range of much larger $x$ values. We need to take
this large-$x$ behaviour uncertainty into account. We do so by letting the
$\beta$ parameters vary; we consider a grid of configurations in which
$\beta_u$ and $\beta_d$, both for the transversity and the Sivers distributions,
range from $0$ to $4$ in steps of $0.5$.
For each of these configurations, we re-run the SIDIS best fit.
Then we select out only the parameter
configurations that correspond to a $\chi^2_{dof}$ not larger than about $20\%$
more of the minimum original value~\cite{Anselmino:2008sga, Anselmino:2008jk}.
Finally, we construct a variation band, for both the Sivers and Collins

\begin{wrapfigure}{r}{0.57\columnwidth}
\vspace*{-0.4cm}
\begin{center}
\includegraphics[width=0.45\textwidth, angle=-90]{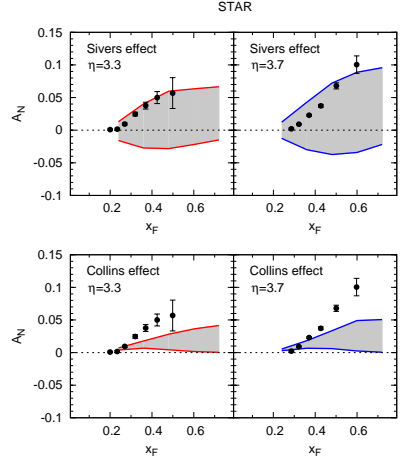}
\caption{The Sivers (upper panel) and Collins (lower panel) effects, evaluated
according to Eq.~(\ref{AN}) by using the Sivers, transversity and Collins
functions extracted in~\cite{Anselmino:2008sga,Anselmino:2008jk}, are compared
to the experimental data from STAR~\cite{star}. The shaded bands are obtained by
scanning over the $\beta$ parameters, as described in the text.}
\label{Fig:Sivers-effect}
\end{center}
\end{wrapfigure}
%
\noindent
contributions to $A_N$, given by the convolution of all the curves obtained
from the parameter sets we have selected.

\begin{wrapfigure}{l}{0.57\columnwidth}
\vspace*{-0.9cm}
\begin{center}
\includegraphics[width=0.45\textwidth, angle=-90]{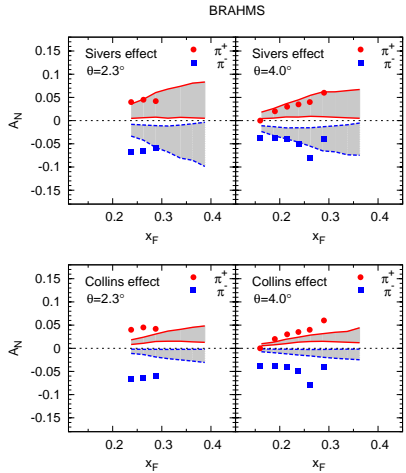}
\end{center}
\vspace*{-0.6cm}
\caption{The Sivers (upper panel) and Collins (lower panel) effects, evaluated
according to Eq.~(\ref{AN}) by using the Sivers, transversity and Collins
functions extracted in~\cite{Anselmino:2008sga,Anselmino:2008jk}, are compared
to the experimental data from BRAHMS~\cite{brahms}.
}
\label{Fig:Collins-effect}
\end{wrapfigure}

\noindent
Figures~\ref{Fig:Sivers-effect} and ~\ref{Fig:Collins-effect} show the results
we obtain, from which we can draw a few conclusions. We do not find any clear
strong indication of universality breaking effects; on the contrary, given
the constraints offered by the presently available SIDIS and $e^+e^-$
experimental data,
our results show that there might exist a set of SIDIS
extracted Sivers functions which can account for the transverse single spin
asymmetry $A_N$ for neutral and charged pion production in polarized
proton-proton scattering measured by RHIC~\cite{star,brahms}.
Instead, the Collins effect alone, for which universality is usually accepted,
only contributes a fraction of the whole $pp$ asymmetries.

~\\
~\\~

%
\noindent
The study of the dependence of these results on the choice of the fragmentation
function set is currently under way. 
There is evidence that this dependence,
which is quite mild in SIDIS processes, can be much more pronounced in the case
of $pp$ scattering where the cross sections become much more sensitive to
the details of the gluon distribution function (recall that there is no glue contribution to SIDIS processes at LO).
Moreover the $Q^2$ evolution of the
Sivers and of the Collins functions are yet unknown.
\noindent
In our fit we assume the
same evolution as that of the corresponding unpolarized density functions:
the consequences of this simplification have to be analyzed in more details.

\noindent
These results are entirely phenomenological and preliminary: further studies
and data are obviously necessary before one can definitely conclude whether
or not the same sets of Sivers and Collins distributions, within a TMD
factorized scheme, can explain the SSAs measured in SIDIS and hadronic processes.
At the moment we can only conclude that, within the large variation bands,
the sum of the (SIDIS extracted) Sivers and Collins contributions could
fit the RHIC data on $A_N$ at $\sqrt s = 200$ GeV.

~~~

%
%
%

\begin{footnotesize}

\end{footnotesize}


\end{document}